\documentclass[runningheads]{llncs}
\usepackage{booktabs}
\usepackage[T1]{fontenc}
\usepackage[utf8]{inputenc}
\usepackage{graphicx}
\usepackage{listings}
\usepackage{hyperref}
\usepackage{url}
\usepackage{todonotes}
\usepackage{wrapfig}
\usepackage{lipsum}
\usepackage{tikz}
\usetikzlibrary{arrows.meta, positioning, shapes.geometric, calc}
\usepackage{amsfonts}

\pdfoutput=1 

\begin{document}

\title{Hammering Higher Order Set Theory}
\titlerunning{Hammering Higher Order Set Theory} 
\author{Chad E. Brown\inst{1} \and
  Cezary Kaliszyk\inst{2}\orcidID{0000-0002-8273-6059} \and
    Martin Suda\inst{1} \and
  Josef Urban\inst{1}\orcidID{0000-0002-1384-1613}
}
\authorrunning{C. Brown, C. Kaliszyk, M. Suda, J. Urban}
\institute{Czech Technical University in Prague, Czech Republic\\
\email{martin.suda@gmail.com,josef.urban@gmail.com} \and
  University of Melbourne, Australia and University of Innsbruck, Austria\\
\email{ckaliszyk@unimelb.edu.au}}
\maketitle

\lstdefinelanguage{megalodon}{
  keywords={Theorem, apply, rewrite, let, assume, exact, Qed, fun, hammer, claim},
  keywordstyle=\color{blue}\bfseries,
  sensitive=true,
  morekeywords={/\,, \/\, ~, ->, =},
  commentstyle=\color{gray}\itshape,
  morecomment=[l]{//},
  morestring=[b]",
  morestring=[b]',
  basicstyle=\ttfamily\small
}

\lstset{
  language=megalodon,
  basicstyle=\ttfamily, 
  keywordstyle=\color{blue}\bfseries,
  commentstyle=\color{gray},
  stringstyle=\color{red},
  numbers=none, 
  numberstyle=\tiny\color{gray},
  stepnumber=2,
  numbersep=10pt,
  showstringspaces=false,
  backgroundcolor=\color{white},
  breaklines=true,
  breakatwhitespace=true,
  columns=flexible,
  frame=single,
  tabsize=2,
  literate=
    { /\\ }{{\color{violet}$\land\;$}}1
    { \\/ }{{\color{violet}$\lor\;$}}1
    { - > }{{\color{violet}$\rightarrow\;$}}1
    { ~ }{{\color{violet}$\neg$}}1}


\begin{abstract}
  We use automated theorem provers to significantly shorten
  a formal development in higher order set theory.
  The development includes many standard theorems
  such as the fundamental theorem of arithmetic and
  irrationality of square root of two.
  Higher order automated theorem provers are particularly
  useful here, since the underlying framework of higher order set theory
  coincides with the classical extensional higher order logic of
  (most) higher order automated theorem provers,
  so no significant translation or encoding is required.
  Additionally, many subgoals are first order
  and so first order automated provers often suffice.
  We compare the performance of different provers
  on the subgoals generated from the development.
  We also discuss possibilities for proof reconstruction,
  i.e., obtaining formal proof terms
  when an automated theorem prover claims to have proven the subgoal.
\end{abstract}

\section{Introduction}\label{s:intro}


In this work, we first describe (Section~\ref{s:devel}) a formal
development in higher order set theory (using the Megalodon system)
containing several well-known mathematical theorems, including
Conway’s surreal numbers and 12 of the Freek100 theorems.  We then use
this development (Section~\ref{s:atp}) to create a large set (and a
benchmark) of higher order 
problems for
automated theorem provers (ATPs), evaluating multiple state-of-the-art
ATPs on the benchmark.  When an ATP is able to solve one of the
problems, we are able to replace part of the development with a call
to an ATP.  This results in a reduction of the text of the development
to less than a half its original size. We also describe the related
hammer-style proof automation for Megalodon in Emacs, and discuss
proof reconstruction (Section~\ref{s:reconstruction}).



\section{Megalodon and Formalization of 12 Freek100 Theorems}\label{s:devel}

\def\limplies{\Rightarrow}

\subsection{Megalodon}
We start with a formal development in the Megalodon system.
Megalodon is a fork of the Egal system~\cite{BrownPak19}
and is based on higher order Tarski-Grothendieck set theory.
The underlying logical framework is simply-typed intuitionistic higher order logic
(with Curry Howard proof terms).
In addition to the built-in type $o$ of propositions
we use one base type $\iota$ which should be interpreted as sets.
The remaining types are function types $\alpha\to\beta$, as usual.
Simply typed terms are constructed via the grammar
$$c \mid x \mid (s\, t) \mid (\lambda x.t) \mid (s\limplies t) \mid (\forall x.t)$$
where $c$ ranges over (typed) names (which may be primitives or definitions)
and
$x$ ranges over (typed) variables.
We assume the usual conventions about simple type theory (e.g., $s\, t\, u$ means $((s\, t)u)$
and $\lambda x y.t$ means $\lambda x.\lambda y.t$)
and leave the reader to consult~\cite{BrownPak19} if necessary.
Although we have only included implication and universal quantification,
there are (several) well-known impredicative definitions of other connectives and quantifiers~\cite{Russell03,Prawitz06,Brown2015}.
Consequently, we will freely use $\top$, $\bot$, $\lnot$, $\land$, $\lor$, $\Leftrightarrow$,
$\exists$ and $=$ below without further comment.

There are well-known ways to axiomatize set theory within a simply typed framework~\cite{Paulson93,Gordon96}.
In our case we add
a primitive $\varepsilon$ as a choice function over $\iota$
and primitives of the set theory: $\in:\iota\to\iota\to o$, $\emptyset:\iota$, $\bigcup:\iota\to\iota$ (unary union operator),
$\wp:\iota\to\iota$ (power set operator), ${\tt{Repl}}:\iota\to (\iota\to\iota)\to \iota$ (Fraenkel Replacement operator) and ${\tt{UnivOf}}:\iota\to\iota$.
The unary operator ${\tt{UnivOf}}$ takes a set $X$ and returns the least Grothendieck universe
with $X$ as a member.
A Grothendieck universe is a transitive set
closed under $\bigcup$, $\wp$ and ${\tt{Repl}}$.
Typically a Grothendieck universe is also required to have an infinite set,
so that it would be a model of Zermelo Fraenkel. Here we also consider $V_\omega$ (the
set of all hereditarily finite sets) to be a Grothendieck universe.
Applying ${\tt{UnivOf}}$ to the empty set yields the set of hereditarily finite sets.
Since $V_\omega$ is infinite, we do not need an explicit axiom of infinity.
In addition to the primitives, we assume the expected set theoretic axioms,
including set extensionality. We also assume a higher order axiom of $\in$-induction
from which Regularity can be proven.
Finally we assume axioms of propositional and functional extensionality so that
the underlying higher order logic is extensional.
The full set of primitives and axioms can be found in~\cite{BrownPak19},
with the minor change that here we only assume $\varepsilon$ at type $\iota$
rather than at every type.
\subsection{Initial Formalization of 12 Freek100 Theorems}
As one would expect, the development continues by making definitions and proving theorems.
Wiedijk maintains a collection of 100 
theorems
that have become standard challenges for ITPs.\footnote{\url{https://www.cs.ru.nl/~freek/100/}}
For our purposes we limit the development to definitions and theorems
that are sufficient to prove 12 of these 100 theorems, mostly removing definitions
or lemmas that are not dependencies of at least one of these 12, or a variant of one of the 12.
Listed in order of appearance in the development,
the selected 12 theorems are mathematical induction,
Cantor's Theorem, Schroeder Bernstein, number of subsets of a (finite) set,
infinitude of primes, Ramsey's Theorem, Bezout's Theorem, the greatest common divisor algorithm,
the Fundamental Theorem of Arithmetic, non-denumerability of the reals, denumerability of the rationals
and the irrationality of $\sqrt{2}$. 
In order to state these results, the development must include a representation
of functions, natural numbers, integers, rationals and reals. 

The first part of the development contains basic logical definitions and theorems.
For example, conjunction ($\land$) is defined as $\lambda A B.\forall p.(A\limplies B \limplies p)\limplies p$
and a theorem ${\mathtt{andI}}$ $\forall A B.A\limplies B\limplies A \land B$ is proven
with the ${\mathtt{exact}}$ tactic and the proof term $\lambda A~B~a~b~p~H.H~a~b$.
Later in the development when a subgoal has a conclusion of the form $A\land B$,
we can reduce the goal to the obvious two subgoals ($A$ and $B$) using
the tactic ${\mathtt{apply}}~{\mathtt{andI}}$. This already introduces a minor
amount of automation, as the ${\mathtt{apply}}$ tactic uses matching to extract
the $A$ and $B$ from the subgoal in order to construct the partial proof term
${\mathtt{andI}}~A~B~D~E$ where $D$ and $E$ will be determined by the
subsequent subproofs.

The tenth theorem of the entire development (well before
any of the 12 Freek100 theorems) is excluded middle ${\mathtt{xm}}: \forall p.p\lor\lnot p$.
The proof follows the Diaconescu argument~\cite{Diaconescu75,GoodmanMyhill78}
using $\varepsilon$ at $\iota$. Before proving excluded middle,
it does not make sense to use classical ATPs as hammers.
For this reason, we only start considering the use of ATPs
after the first 10 theorems of the development.

Throughout the development basic infrastructure is included 
(e.g., ordinals, ordered pairs and functions as sets)
as needed, largely following~\cite{Brown2015}, though the setting there was intuitionistic.
As usual, the finite ordinals are used to represent natural numbers
and provide the material to state and prove mathematical induction, both
in its usual form and in the form of complete induction.
The ordinal $\omega$ is defined to be the set of natural numbers,
and we define a set to be finite if it is equipotent to a member of $\omega$.
This basic material is enough to state and prove five more of the 12 theorems.
The development up to the proofs of these first 6 theorems consists of 6649 lines
with 59 definitions and 314 theorems (including the 6).
\subsection{Conway's surreal numbers} The remaining 6 of the 12 require either integers, rationals or real numbers.
We construct these numbers (and more) using a specific
set-theoretic representation of Conway's surreal numbers~\cite{ConwayONAG2001}.
Surreal numbers have also been formalized in Mizar~\cite{DBLP:conf/itp/PakK24}
using representations closer to Conway's description.
Here we use a different representation, for reasons that will become clear.
A surreal number can be uniquely represented by an ordinal length branch on
an arbitrarily large binary tree.
That is, each surreal number can be represented by an ordinal $\alpha$ and a
function $f:\alpha\to\{0,1\}$ where $f(\beta)=0$ indicates the left option at
stage $\beta\in\alpha$ and $f(\beta)=1$ indicates the right option at stage $\beta\in\alpha$.
Instead of representing surreal numbers in precisely this way, we represent
them using a set that remembers the ordinal $\alpha$ and which ordinals $\beta\in\alpha$
would take the left option and which would take the right option.
To represent the left option at stage $\beta$, we define the set $\beta'$ to be $\beta\cup\{\{1\}\}$.
It is easy to see that $\beta'$ is never an ordinal (since the set $\{1\}$ is not transitive).
It is also to see that if $\beta'=\gamma'$ for ordinals $\beta$ and $\gamma$, then $\beta=\gamma$.
We now represent the surreal number given by $\alpha$ and $f:\alpha\to\{0,1\}$
by the set $\{\beta\in\alpha | f(\beta)=1\}\cup\{\beta' | \beta\in\alpha, f(\beta)=0\}$.
It is easy to see that $\alpha$ and $f$ can be recovered from this set.
It is also easy to see that the surreal number given by $\alpha$ and $f:\alpha\to\{0,1\}$
with $f(\beta)=1$ for all $\beta\in\alpha$ is represented simply by the set $\alpha$.
That is, ordinals are already surreal numbers using this representation,
and correspond to the branch of length $\alpha$ that chooses the right option at each step.
Other examples of surreal numbers are $-2$ represented by $\{0',1'\}$
and $\frac{1}{2}$ represented by $\{0,1'\}$.
One can define recursion operators on surreal numbers and use these
to define the basic operations as described in~\cite{ConwayONAG2001}
essentially proving the surreal numbers form an ordered field (with a proper class as its carrier).
The development up to this point consists of 27316 lines with 116 definitions and 759 theorems,
so that the primary surreal number part of the development consists of
20668 lines with 57 definitions and 445 theorems (almost half the development).
In the next 2544 lines (with 6 new definitions and 57 new theorems),
the set ${\mathtt{Z}}$ of integers can easily be defined (carved from the surreal number field)
allowing us to state and prove Bezout's Theorem, the greatest common divisor algorithm
and the Fundamental Theorem of Arithmetic.
The next 11211 lines (with 11 new definitions and 137 new theorems)
define the set ${\mathtt{R}}$ of real numbers (carved out of the surreals), proven to have the
expected properties including uncountability.
With 730 more lines, 2 new definitions and 8 new theorems,
we can define the set ${\mathtt{Q}}$ of rational numbers
as a subset of ${\mathtt{R}}$ and prove ${\mathtt{Q}}$ is countable.
Note that by construction we have $\omega\subseteq {\mathtt{Z}} \subseteq {\mathtt{Q}}\subseteq {\mathtt{R}}$,
which was our motivation for the specific representation of surreal numbers as sets.

The only remaining theorem of the 12 is irrationality of $\sqrt{2}$.
Conway describes how to recursively define the square root operation on nonnegative surreal numbers.
We make this definition and prove the square root of nonnegative reals are nonnegative reals.
The development ends with the proof that $\sqrt{2}$ is irrational, i.e., $\sqrt{2}\in{\mathtt{R}}\setminus{\mathtt{Q}}$.
This part of the development used 3203 more lines with 4 new definitions and 38 new theorems.

In total the development contains 139 definitions and 999 theorems.
The development is given as a single file with 45004 lines and 346152 characters,
and the proofs are given in significant detail, with little automation.

\section{Development using ATPs}\label{s:atp}
\subsection{Automation Tactic}
We have extended Megalodon to allow the option of using
the tactic ${\mathtt{aby}}$ (``automated by'') with a list of dependencies
to justify a subgoal.\footnote{Megalodon with the data/code discussed here is available at {\url{https://github.com/MgUser36/megalodon}}.}
Megalodon can check the file and
produce problem files for ATPs.
Since the underlying framework is simply typed higher order logic, 
we can directly translate the given dependencies and current subgoal
into the TH0 TPTP format~\cite{THFcite}, a common format for
higher order ATPs.
Hence every use of the ${\mathtt{aby}}$ tactic will result in one TH0 problem file.
In addition, if the conclusion and the dependencies are in
the first order fragment of higher order logic, we can
translate the problem to the FOF TPTP format and call first order theorem provers.
Note that we do not attempt to encode higher order logic into first order \cite{BlanchetteB0S16,Paulson10,hurd2003d,ckju-jar14}.
The use of the ${\mathtt{aby}}$ tactic is justified if either a higher order ATP
can prove the TH0 file, or if a first order ATP can prove the FOF file.

\subsection{Experiments}
In order to identify subproofs that could be replaced by calls to ATPs,
we used Megalodon to generate TH0 problem files for most
tactic calls, determining the dependencies when the subproof is completed.
Each problem corresponds to a sequence of text (from before the tactic is called to where
the subproof is completed) that can be replaced if the problem file can be proven
by an ATP. This resulted in 41738 higher order problem files.\footnote{The resulting large HO-TP benchmark is at \url{https://github.com/MgUser36/MegalodonATPBenchmark}.}
With a 60 second timeout,
we ran Vampire~\cite{VampireHO} (with two different portfolios), Zipperposition~\cite{Zipperposition},
E~\cite{EHO}, Lash~\cite{Lash}
and cvc5~\cite{cvc5}.
The results are shown in Table~\ref{tab:bushy}.
In total, 34172 (82\%) of the 41738
subgoals could be replaced by ATP calls.
Megalodon additionally produced 29880 first order problem files
when the subgoal had a first order proof.
We ran Vampire for 5 seconds on each of these problems and 60 seconds
on a selection of the problems,
resulting in 11179 subproofs that could be replaced by ATP calls,
although only 11 of these subproofs were not already identified by the higher order problems.

\begin{table}
  \begin{center}
  \begin{tabular}{cccccc}
    Vampire & Vampire & Zipperposition & E & Lash & cvc5 \\
    (sledgehammer) & (ho) & & & \\\hline
    32675 & 32474 & 31310 & 23866\,\, &\,\, 14987\,\, &\,\, 13238\,\, \\
    78.3\% & 77.8\% & 75\% & 57.2\% & 35.9\% & 31.7\%\\\\
  \end{tabular}
  \end{center}
  \caption{Higher order ATPs on premise-selected subgoals}\label{tab:bushy}
\end{table}

Some ATP calls may result in replacing a larger part of text than others.
For example, a subproof in the original development contained a line with three
tactics:
\\
\begin{lstlisting}
  apply In_irref delta. rewrite H2 at 2. exact Ldsa.
\end{lstlisting}
This line resulted in three problem files, corresponding to whether
the ${\mathtt{exact}}$ could be replaced, the ${\mathtt{rewrite}}$ and the ${\mathtt{exact}}$ could
be replaced, and whether all three could be replaced.
In this case, all three could be replaced.
That is, one problem solved by an ATP would allow us to
replace the ${\mathtt{exact}}$ tactic by an ${\mathtt{aby}}$ call.
A second problem solved by an ATP would allow us
to replace both the ${\mathtt{rewrite}}$ and the ${\mathtt{exact}}$ tactic
in the line above with an ${\mathtt{aby}}$ call.
A third problem solved by an ATP would allow us
to replace all three tactics on the line by an ${\mathtt{aby}}$ call.
Obviously replacing all three tactics supersedes replacing either the
last two or the last one alone, so it makes sense to find the
ATP problems that allow us to replace as much text as possible.
After filtering for the problems that allow the maximum amount of text to be replaced,
the first two ATP problems above (corresponding to replacing ${\mathtt{exact}}$
alone or only replacing ${\mathtt{rewrite}}$ and ${\mathtt{exact}}$) would
be filtered out, in favor of only including the third ATP problem allowing the
replacement of all three tactics.

Once we restrict to the problems that allow the largest texts to be
replaced (along with some manual modifications),
we were left with 3401 calls to an ATP
and a development with 17435 lines and 159363 characters.
This is roughly 46\% the original size of the development.\footnote{Note that we do not include the ATP output as part of the size of the new text.}
The majority of the proofs in the development (765 of the 989
proofs considered for replacement) could be replaced simply
by a single ${\mathtt{aby}}$ call. The remaining 224 proofs contained
2636 ATP calls in subproofs, with approximately 12 ATP
calls per proof.

In order to justify the 3401 ATP calls, Megalodon produced
3401 TH0 problem files and 1618 FOF problem files.
Table~\ref{tab:hammeratpresults}
shows the results of higher order ATPs on these files.
We also called a recent first order version of Vampire was called on the FOF problems,
with it solving 1322 (81.7\%) of the 1618 first order problems.
Each call was with a 60s timeout.

\begin{table}
  \begin{center}
  \begin{tabular}{cccccc}
    Vampire & Vampire & Zipperposition & E & Lash \\
    (sledgehammer) & (ho) & & & \\\hline
    3223 (94.8\%) & 3165 (93.1\%) & 2801 (82.4\%) & 2403 (70.7\%) & 1567 (46.1\%)\\\\
  \end{tabular}
  \end{center}
  \caption{Higher order ATPs on the 3401 ATP problems}\label{tab:hammeratpresults}
\end{table}

\subsection{Examples}

We consider a few examples to give an idea of what kinds of proofs can be replaced by calls
to an ATP.\footnote{Successful ATP outputs with TPTP proofs for these examples can be found at {\url{http://grid01.ciirc.cvut.cz/~chad/atppfs2025}}.}

\paragraph{Example 1}
The largest text replaced by a call to an ATP (in this case, Lash) is the proof of the following theorem \texttt{PNoLt\_tra}\footnote{\url{https://mgwiki.github.io/mgw_test/Part2.mg.html\#PNoLt_tra}}:\\
\begin{lstlisting}
Theorem PNoLt_tra : forall alpha beta gamma,
   ordinal alpha -> ordinal beta -> ordinal gamma ->
   forall p q r:set -> prop,
     PNoLt alpha p beta q -> PNoLt beta q gamma r
       -> PNoLt alpha p gamma r.
\end{lstlisting}
This is essentially the proof of transitivity of the (strict) ordering on surreal numbers.
At the point in the development where this theorem is stated and proven, surreal numbers
are not yet defined. However, the ordinal $\alpha$ with the predicate $p:\iota\to o$
determines a surreal number where $p$ indicates for which $\alpha'\in\alpha$ the right option
is taken. Likewise, $\beta$ with $q$ determines a surreal number and $\gamma$ with $r$ determines
a surreal number.
The definition of ${\mathtt{PNoLt}}~\alpha~p~\beta~q$ corresponds to
when the surreal number determined by $\alpha$ and $p$ is less than the surreal number
determined by $\beta$ and $q$.
This can happen in three ways:
\begin{enumerate}
\item $\alpha = \beta$ and there is some $\delta\in \alpha$ such that
  $p$ and $q$ agree up to $\delta$,
  $\lnot p~\delta$ and $q~\delta$.
  That is, the two surreal numbers use the same left and right options
  until $\delta$ at which point $p$ chooses the left option and $q$ chooses the right option.
\item $\alpha \in \beta$, $p$ and $q$ agree up to $\alpha$ and $q~\alpha$.
  That is, the surreal number given by $\beta$ and $q$ is an extension of the sequence
  given by $\alpha$ and $p$, where the first new option (at step $\alpha$) is the right option.
\item $\beta \in \alpha$, $p$ and $q$ agree up to $\beta$ and $\lnot p~\beta$.
  That is, the surreal number given by $\alpha$ and $p$ is an extension of the sequence
  given by $\beta$ and $q$, where the first new option (at step $\beta$) is the left option.
\end{enumerate}
To give concrete examples, suppose $\alpha = 1$ and $p~0$ holds.
Then $\alpha$ and $p$ correspond to the surreal number $1$, since the sequence
is of length $1$ and chooses the right option at step $0$.
Suppose $\beta = 2$ and both $q~0$ and $q~1$ hold.
Then $\beta$ and $q$ correspond to the surreal number $2$.
In this case ${\mathtt{PNoLt}}~\alpha~p~\beta~q$ holds since $\alpha\in\beta$,
$p$ and $q$ agree up to $1$ (i.e., $p~0 \Leftrightarrow q~0$)
and $q~1$ holds.
On the other hand, if $\beta = 2$, $q~0$ and $\lnot q~1$ (corresponding to the
surreal number $\frac{1}{2}$)
then we have ${\mathtt{PNoLt}}~\beta~q~\alpha~p$ holds since $\alpha\in\beta$,
$p$ and $q$ agree up to $1$ and $\lnot q~1$ holds.
Finally, suppose $\alpha = 2$, $p~0$, $\lnot p~1$,
$\beta = 2$, $q~0$ and $q~1$,
then ${\mathtt{PNoLt}}~\alpha~p~\beta~q$ holds since
$\alpha = \beta$, $p$ and $q$ agree up to $1$, $\lnot p~1$ and $q~1$.

The manual proof of transitivity of ${\mathtt{PNoLt}}$ proceeds as follows.
Suppose $\alpha$, $\beta$ and $\gamma$ are ordinals and $p,q,r:\iota\to o$
are such that
${\mathtt{PNoLt}}~\alpha~p~\beta~q$
and ${\mathtt{PNoLt}}~\beta~q~\gamma~r$ hold.
We need to prove ${\mathtt{PNoLt}}~\alpha~p~\gamma~r$ holds.
We split into the three cases based on why ${\mathtt{PNoLt}}~\alpha~p~\beta~q$ holds.
In each of the three cases we split into three subcases
based on why ${\mathtt{PNoLt}}~\beta~q~\gamma~r$ holds.
The full manual proof uses 311 lines. Of these 311 lines,
the first 5 are simply introducing the variables and hypotheses.
Roughly 61 lines correspond to splitting into the 3 cases and the 9 subcases,
introducing the relevant variables and hypotheses for each case and subcase.
On average completing the proof in each of the 9 subcases requires about 27 lines.
When using the hammer, the full proof is reduced to the following single use of ${\mathtt{aby}}$:\vspace{1cm}
\begin{lstlisting}
aby and3I binintersectI binintersectE ordinal_Hered ordinal_trichotomy_or PNoEq_tra_ PNoEq_antimon_ PNoLtI1 PNoLtI2 PNoLtI3 PNoLtE.
\end{lstlisting}
While the automated proof is much shorter, a reader examining the proof
may have trouble determining that the proof involves splitting into 3 cases and
9 subcases. The only hint is that the call to ${\mathtt{aby}}$ references ${\mathtt{PNoLtE}}$,
which is a theorem that can be applied when we know ${\mathtt{PNoLt}}~\alpha~p~\beta~q$
to split the current goal into three cases.

\paragraph{Example 2}
The second largest text replaced in a proof is at the end of the
proof of the following theorem \texttt{PNo\_rel\_split\_imv\_imp\_strict\_imv}\footnote{\url{https://mgwiki.github.io/mgw_test/Part2.mg.html\#PNo_rel_split_imv_imp_strict_imv}}:
\begin{lstlisting}
Theorem PNo_rel_split_imv_imp_strict_imv :
 forall L R:set -> (set -> prop) -> prop,
  forall alpha, ordinal alpha -> forall p:set -> prop,
       PNo_rel_strict_split_imv L R alpha p
    -> PNo_strict_imv L R alpha p.
\end{lstlisting}
Again, this theorem is about surreal numbers before committing to the particular set theoretic
representation of surreal numbers.
Here $L$ and $R$ of type $\iota\to (\iota\to o) \to o$ can be thought of as sets of
surreal numbers (given as $\beta$ and $q:\iota\to o$).
The conclusion ${\mathtt{PNo\_strict\_imv}}~L~R~\alpha~p$ means
that the surreal number given by $\alpha$ and $p$ is an intermediate value between $L$ and $R$.
That is, ${\mathtt{PNoLt}}~\beta~q~\alpha~p$ whenever $L~\beta~q$
and ${\mathtt{PNoLt}}~\alpha~p~\beta~q$ whenever $R~\beta~q$.
The hypothesis ${\mathtt{PNo\_rel\_strict\_split\_imv}}~L~R~\alpha~p$
is a similar property about the two extensions of $\alpha$ and $p$ of length $\alpha^+$.

The manual proof of this theorem was 240 lines.
The proof using ${\mathtt{aby}}$ is 27 lines.
No ATP was able to prove the theorem completely. (If an ATP had
proven the full theorem, we could
replace the full proof by one call to ${\mathtt{aby}}$.)
The proof using ${\mathtt{aby}}$ starts the same way as the manual proof,
assuming we have $L$, $R$, an ordinal $\alpha$ and some $p:\iota\to o$ satisfying
${\mathtt{PNo\_rel\_strict\_split\_imv}}~L~R~\alpha~p$.
ATPs can automatically prove $\alpha^+$ (the successor of $\alpha$) is also an ordinal, so this
subproof is replaced by ${\mathtt{aby}}$.
As part of the beginning of the proof we define $p_0$ to be the predicate
that holds for $\delta$ if $p~\delta\land \delta\not=\alpha$
and we define $p_1$ to be the predicate
that holds for $\delta$ if $p~\delta\lor \delta=\alpha$.
ATPs can automatically prove $\lnot p_0~\alpha$ and $p_1~\alpha$,
so these two subproofs are replaced by ${\mathtt{aby}}$.
ATPs can also prove
${\mathtt{PNoLt}}~\alpha^+~p_0~\alpha~p$
and 
${\mathtt{PNoLt}}~\alpha~p~\alpha^+~p_1$,
so these two subproofs are replaced by ${\mathtt{aby}}$.
At this point we need to prove the conclusion
${\mathtt{PNo\_strict\_imv}}~L~R~\alpha~p$.
The rest of the manual proof requires 200 lines.
ATPs can finish the rest of the proof and so these 200 lines
are replaced by a final call to ${\mathtt{aby}}$.
In the manual proof, the first of the remaining 200 lines
explicitly expands the definition of ${\mathtt{PNo\_strict\_imv}}$
and reduces the conclusion to proving $\alpha$ and $p$ are greater
than everything in $L$ and to proving $\alpha$ and $p$ are less
than everything in $R$.
Meanwhile, when these last 200 lines are replaced by an ${\mathtt{aby}}$
call, there is no clear indication that the reader should expand
the definition of ${\mathtt{PNo\_strict\_imv}}$ and check the two conjuncts.
This is another example where replacing subproofs by ${\mathtt{aby}}$ may make
the proof less readable by making it less explicit.

\paragraph{Example 3}
In many cases, replacing details of a proof may make it easier to read a proof.
Consider the following theorem \texttt{exp\_SNo\_nat\_mul\_add}\footnote{\url{https://mgwiki.github.io/mgw_test/Part6.mg.html\#exp_SNo_nat_mul_add}}:
\begin{lstlisting}
Theorem exp_SNo_nat_mul_add : forall x, SNo x ->
  forall m, nat_p m -> forall n, nat_p n ->
     x ^ m * x ^ n = x ^ (m + n).
\end{lstlisting}
Here $x^n$ is the operation raising a surreal number $x$ to the $n^{th}$ power (for a natural number $n$).
The goal is to prove $x^m\cdot x^n = x^{m+n}$, where multiplication and addition are over surreal numbers
(which agree with multiplication and addition on natural numbers).
The manual proof fixes $x$ and $m$ and proves that $m$ is a surreal number (since it is natural).
We then apply natural number induction, reducing the problem to proving the case for $n=0$
and the inductive case.
At this point in the manual proof a series of explicit (though obvious) arithmetical steps
are required -- including steps to change between addition on natural numbers and addition on
surreal numbers. Meanwhile, ATPs can complete the proofs for the base case and inductive case,
resulting in the following shorter proof:
\begin{lstlisting}
let x. assume Hx. let m. assume Hm.
claim Lm: SNo m.
{ aby nat_p_SNo Hm. }
apply nat_ind.
- aby add_SNo_0R mul_SNo_oneR exp_SNo_nat_0 SNo_exp_SNo_nat
      Lm Hm Hx.
- aby add_nat_SR add_nat_p nat_p_omega omega_ordsucc
      add_nat_add_SNo mul_SNo_com mul_SNo_assoc exp_SNo_nat_S
      SNo_exp_SNo_nat Hm Hx.
Qed.
\end{lstlisting}
The original manual proof is longer (29 lines instead of 6 lines) and all the
extra details are simply distracting.
Omitting these steps arguably makes the proof easier to read.

\subsection{Use in Emacs and a Full Hammer} 
\begin{figure}[t!]
  \centerline{\includegraphics[width=\textwidth]{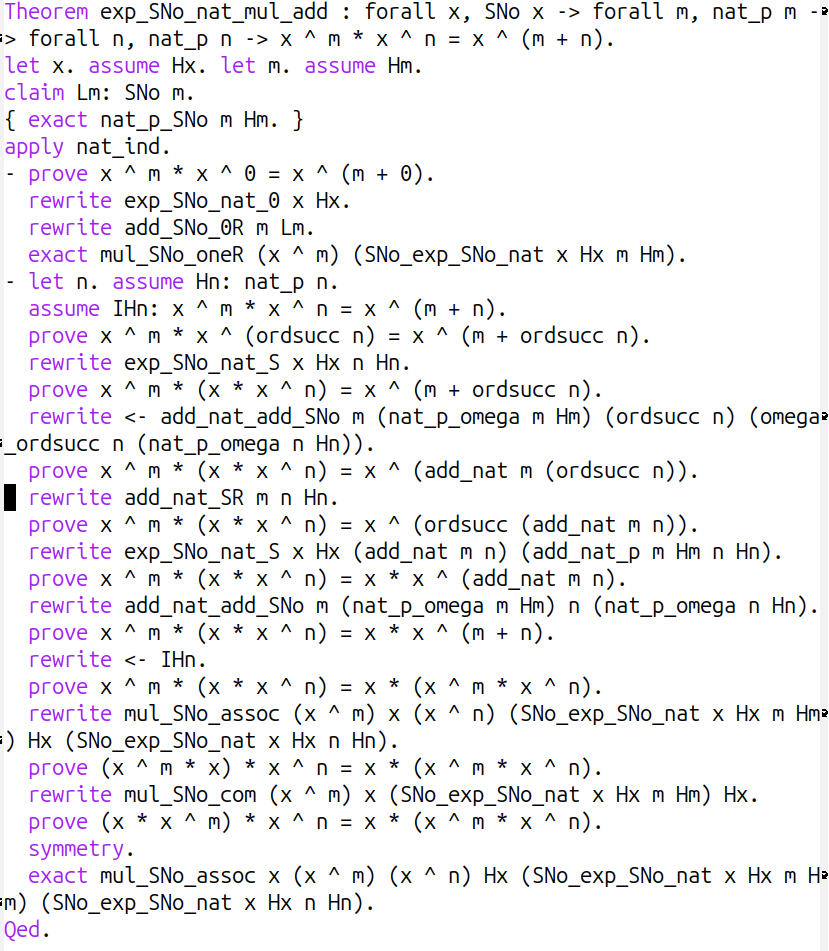}}
  \caption{\label{fig:adv1}Original Megalodon Proof.}
\end{figure}

\begin{figure}[t!]
  \centerline{\includegraphics[width=\textwidth]{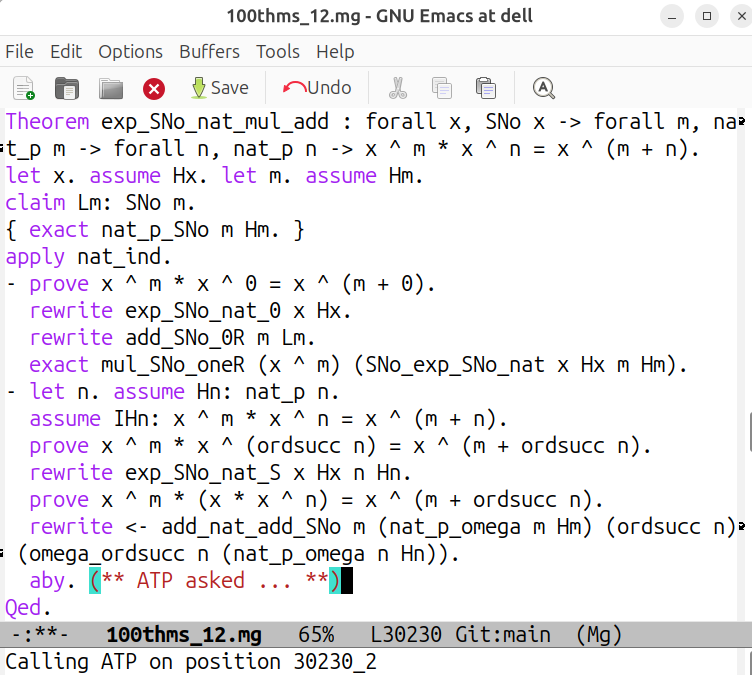}}
  \caption{\label{fig:adv2}Aby call invoked inside the proof.}
\end{figure}

\begin{figure}[t!]
  \centerline{\includegraphics[width=\textwidth]{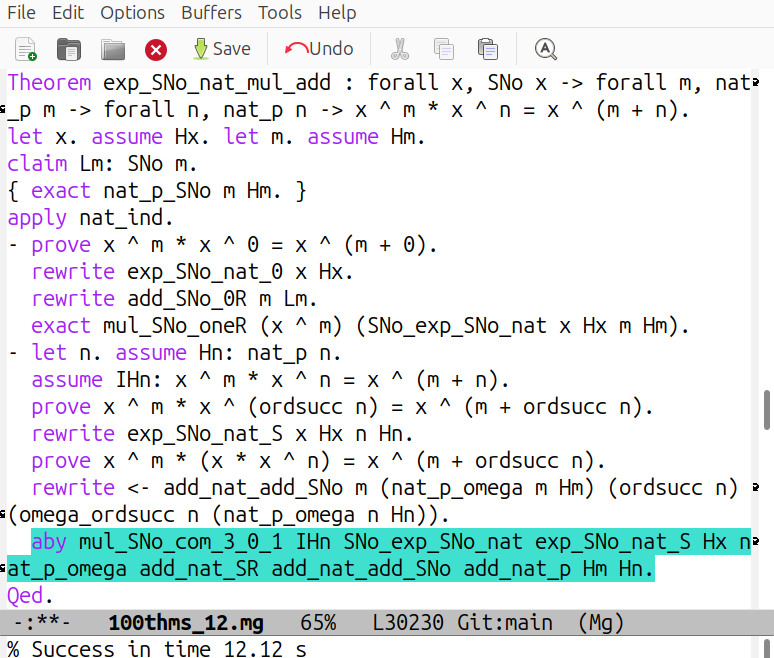}}
  \caption{\label{fig:adv3}Successful aby call with the axiom names inserted.}
\end{figure}

We have also implemented a simple Emacs
mode\footnote{\url{https://github.com/MgUser36/megalodon/blob/main/mg-advice.el}}
for Megalodon and a command in it (${\mathtt{aby.}}$) which implements the ``hammer''
behavior, similar to the Sledgehammer command in
Isabelle/JEdit. The command calls Megalodon to create a TH0 file which
includes the whole previous development. Then it calls Vampire with a
schedule constructed for the Sledgehammer division of the CASC competition.\footnote{The schedule was constructed according to {\url{https://tptp.org/CASC/J11/SystemDescriptions.html\#SnakeForV4.7---1.0}}}
 If successful, the names of the
used axioms are translated back to the Megalodon syntax and the
${\mathtt{aby}}$ call using them is automatically inserted into the current buffer.
The operation is shown in Figures~\ref{fig:adv1}, \ref{fig:adv2}, and \ref{fig:adv3}.

To test this ``full hammer'' performance, we generated corresponding ATP problems each time a tactic was used in the
original development.
This resulted in 51243 TH0 problems.\footnote{Also in our benchmark: \url{https://github.com/MgUser36/MegalodonATPBenchmark/tree/main/chainy}.}
Note that this is larger than the
41738 premise selected problem files.
This difference arises from several factors, which we do not discuss in detail here.
We called Vampire with the two portfolios for 60s.
With the sledgehammer schedule, Vampire proved 29256 problems (57.1\%).
With the default higher-order (ho) schedule, Vampire proved 28811 problems (56.2\%).
The union of the two sets was 30473 problems (59.5\%).



\section{Proof Reconstruction}\label{s:reconstruction}

With the original development, complete proof terms are constructed by
Megalodon from the given proof scripts.
Once ATPs are used,
Megalodon is essentially behaving like the Naproche system~\cite{Naproche}.
That is, when an ATP completes the proof of a subgoal,
we trust that there is, indeed, a proof.
Ultimately it would be more satisfying to do {\emph{proof reconstruction}}
so that we still obtain complete proof terms even when some subgoals are
completed by ATPs.

A common technique~\cite{MENG200941}
is to use an ATP to prune the dependencies
and then use an internal search procedure (like Metis~\cite{hurd2003d} for Isabelle)
to create the appropriate ITP proof. 
Megalodon currently has no significant internal search procedure
that could be used for this purpose.
A more optimal state of affairs would be if ATPs returned
sufficiently detailed proof objects that could be independently checked.
Prover9~\cite{prover9-mace4} is a rare example of an ATP
that has provided explicit proof objects for decades, though it is limited to clausal proofs. This is used for proof reconstruction
e.g. in the Metamath hammer~\cite{DBLP:conf/itp/CarneiroBU23}.

\subsection{Using Vampire with the Deducti Output}
Due to recent work on Vampire~\cite{komel2025casestudyverifiedvampire} 
one can request proof output (for first order problems)
that can be checked by Dedukti~\cite{dedukti},
with some limitations noted below.

We consider a very simple example.
Suppose we are in a proof where we have a set $\alpha$
and a local assumption $Ha$ that $\alpha$ is an ordinal (i.e., the proposition ${\mathtt{ordinal}}~\alpha$).
We have a local subgoal to prove $\alpha+$ is an ordinal (i.e., ${\mathtt{ordinal}}~({\mathtt{ordsucc}}~\alpha)$).
The relevant previous theorem is ${\mathtt{ordinal\_ordsucc}}$:
$\forall \alpha.{\mathtt{ordinal}}~\alpha\limplies {\mathtt{ordinal}}~({\mathtt{ordsucc}}~\alpha)$.
This is 
a first order subgoal and Vampire can easily prove it.
The simplest Megalodon proof term for this subgoal would be
${\mathtt{ordinal\_ordsucc}}~\alpha~Ha$. In Megalodon this part of the proof\footnote{\url{https://github.com/MgUser36/megalodon/blob/main/examples/form100/100thms_12.mg\#L9964}}
looks as follows:
\\
\begin{lstlisting}
claim Lsa: ordinal (ordsucc alpha).
{ exact ordinal_ordsucc alpha Ha. }
\end{lstlisting}

Replacing the subproof with a call to an ATP simply changes the proof to\footnote{\url{https://github.com/MgUser36/megalodon/blob/main/examples/hammer/100thms_12_h.mg\#L3681}}
\\
\begin{lstlisting}
claim Lsa: ordinal (ordsucc alpha).
{ aby ordinal_ordsucc Ha. }
\end{lstlisting}

We examine parts of the Dedukti output of Vampire.
The previous theorem ${\mathtt{ordinal\_ordsucc}}$ is given as an axiom
(with some changes to the name for TPTP compliance) and this is reflected
in the following declared axiom in the Dedukti output:

\begin{verbatim}
{|axiom_ordinal_5Fordsucc9|}:
  Prf (forall iota
           (0 : El iota =>
                 (imp ({|ordinal|} 0)
                      ({|ordinal|} ({|ordsucc|} 0))))).
\end{verbatim}

Likewise the local assumption that $\alpha$ is an ordinal is reflected in the following declared axiom:

\begin{verbatim}
{|axiom_c_Ha16|}: Prf ({|ordinal|} {|alpha|}).
\end{verbatim}

The negated conclusion is also given as a declared axiom:

\begin{verbatim}
{|axiom_18|}: Prf (not ({|ordinal|} ({|ordsucc|} {|alpha|}))).
\end{verbatim}

In principle the rest of the proof should be given as a sequence of Dedukti definitions,
but in practice some of the steps to produce clausal form are left unjustified at the moment.
If we skip forward past the unjustified steps (which are minor in this case),
we can start from three Dedukti items with types corresponding to the three
assumptions above, but in clausal form.
The two resolution steps corresponds to two Dedukti definitions.
In principle the Dedukti representation could be translated into Megalodon as follows:
One begins the subproof by applying double negation, so that one may
assume the negated conclusion and reduce to proving false.
Then each Dedukti definition would correspond to a subgoal
with a proof term translated from its definition.
The last Dedukti definition should give a proof of false (corresponding to the empty clause
in the resolution proof).

This translation 
is left as future work here.
Another general idea for distributing ITP proofs -- which includes proof reconstruction -- is outsourcing
to the (Megalodon-related) Proofgold network~\cite{ProofGold2022}, which newly includes
also the basis for a lightning network~\cite{brown2025payment}.


\section{Related Work}

ITP/ATP hammers and their various components have been developed and
successfully used for over two
decades~\cite{Hurd99,Dahn98,Hurd02,Urb03,Urb04-MPTP0,MENG200941,Paulson10,BlanchetteBP11,BlanchettePhd,holyhammer,hh4h4,BlanchetteGKKU16,DBLP:journals/jar/CzajkaK18,DBLP:conf/itp/CarneiroBU23},
in systems such as Isabelle, HOL, Mizar, HOL Light, Coq, and Metamath. A comprehensive overview of the
topic is given in~\cite{hammers4qed}. Similarly, a number of large
problem sets and benchmarks for higher-order ATPs have been extracted
from various ITP libraries in the recent
years~\cite{DBLP:conf/mkm/BrownU16,DBLP:conf/cade/BrownGKSU19,DBLP:conf/itp/DesharnaisVBW22}.

\section{Conclusion and Future Work}

We have shown that calls to ATPs can
be used to replace over half the content of formal proofs in
a mathematical development in higher order set theory.
The original development (without automation) can be
used to generate tens of thousands of higher order ATP problems
and this benchmark can be used to judge the performance
of modern higher order ATPs, currently showing Vampire as the clear leader.
After replacing as much text as possible with calls to ATPs,
we obtain 3401 higher order ATP problems that are, in some sense,
at the frontier of what current higher order ATPs can solve.

We have also discussed options for 
proof reconstruction, such as the emerging Deducti proof format newly implemented in Vampire, and Megalodon's connection to the Proofgold blockchain, which targets decentralized and distributed proof development.
Another interesting avenue would be the use of SMT solvers,
specifically when the subgoal is about integers or real numbers.
Of course, SMT solvers should only be called once integers (or reals)
are defined (along with the corresponding operations)
and the basic properties have been proven.

\paragraph{Acknowledgments} The results were supported by the Czech
Ministry of Education, Youth and Sports within the dedicated program
ERC CZ under the project POSTMAN no. LL1902, the ERC PoC grant
\emph{FormalWeb3} no. 101156734, Amazon Research Awards, and the Czech
Science Foundation grant no. 25-17929X.

\bibliographystyle{plainurl}
\bibliography{p}

\end{document}